\documentclass[prb, showpacs,twocolumn, preprintnumbers]{revtex4}
\usepackage {graphicx}
\pdfoutput=1

\begin{document}

\title{Effect of self-affine fractal characteristics of surfaces on wetting}

\author{S. Sarkar$^a$, S. Patra$^a$\footnote{Has equal contribution as first author}, N. Gayathri$^b$\footnote{Present address: VECC, 1/AF Bidhannagar, Kolkata 700064}, S. Banerjee$^c$\footnote{Email:sangam.banerjee@saha.ac.in}}

\address{$^a$ Department of Electronics and Communication Engineering, Haldia Insitute of Technology - Haldia West Bengal 721 657 India \\
$^b$ Material Science Division,Indira Gandhi Center for Atomic Research Kalpakkam 603 102 India \\
$^c$ Surface Physics Division,Saha Institute of Nuclear Physics 1/AF Bidhannagar Kolkata 700 064 India \\ }

\begin{abstract}
The relation between the contact angle of a liquid drop and the morphological parameters of self-affine solid surfaces have been investigated. We show experimentally that the wetting property of a solid surface crucially depends on the surface morphological parameters such as: (1) root mean square (rms) roughness $\sigma$, (2) in-plane roughness correlation length $\xi$ and (3) roughness exponent $\alpha$ of the self-affine surface. We have shown that the contact angle monotonically decreases with the increase in the rms local surface slope $\rho$ ($\propto \sigma/\xi^\alpha$) for the cases where the liquid wets the crevices of the surface upon contact. We have shown that the same solid surface can be made hydrophobic or hydrophilic by merely tuning these self-affine surface morphological parameters. 
\end{abstract}

\maketitle

Surface morphology plays a very crucial role in determining the wettability of a surface \cite{Quere,Nosonovsky}. Knowledge about wetting in the sub-microscopic level will have immense impact in narrow channels fluid-flow technology \cite{Extrand,Kusumaatmaja} and in biological applications \cite{Todd}. Wetting on a smooth and flat solid substrate could be well described by Young's \cite{Young} equation and on rough surfaces of solid could be well described by Wenzel's \cite{Wenzel} or Cassie-Baxter's \cite{CB} formalism. Wettability of a surface is mainly determined by three energies: the solid/vapor (SV) interfacial energy $\gamma_{SV}$, solid/liquid (SL) interfacial energy $\gamma_{SL}$ and liquid/vapor (LV) interfacial energy $\gamma_{LV}$. Young pointed out the relationship between the contact angle and the above parameters. The Young's equation is expressed as: $cos\theta_Y = (\gamma_{SV} - \gamma_{SL})/\gamma_{LV}$ or $\gamma_{SL} = \gamma_{SV} - \gamma_{LV}cos\theta_Y$ where $\theta_Y$ is the contact angle the liquid droplet makes with the smooth and flat solid surface. When $\gamma_{SV} > \gamma_{SL}$ then $0^o < \theta_Y < 90^o$ and for $\gamma_{SL} > \gamma_{SV}$ then $90^o < \theta_Y < 180^o$. Introduction of roughness on solid surfaces, increases its interfacial energy with liquid and vapor. Wenzel \cite{Wenzel} proposed that the modified contact angle due to roughness can be expressed as $cos\theta_W = r (\gamma_{SV} - \gamma_{SL})/\gamma_{LV} = r cos\theta_Y$ where $"r"$ is the ratio between the true area $A$ and the projected area $A_0$ ($r=A/A_0$) and is always greater than 1. Cassie and Baxter \cite{CB} further extended this by considering the solid/liquid (SL) interface to have air pockets in the crevices below the liquid drop due to high solid asperities (features like -"Lotus effect"), where some fraction of the solid $f_S$ is in contact with the liquid drop and some fraction $f_V$ under the drop has air or vapor pockets. By including the area ratio factor $"r"$, Cassie and Baxter showed the dependence of the modified contact angle as a function of $f_S$ i.e., $cos\theta_{CB} = rf_S cos\theta_Y + f_S -1$.  Here, as $f_S$ tends to zero, the contact angle approaches 180$^o$ and as $f_S$ tends to 1, the expression tends to the Wenzel's equation. Recently we have shown a simple derivation of Young, Wenzel and Cassie-Baxter equation starting from the work of adhesion \cite{sangam}.

The area ratio factor "$r$" depends mainly on three morphological parameters (1) root mean square (rms) roughness $\sigma$, (2) in-plane roughness correlation length $\xi$ and (3) roughness exponent $\alpha$. If these parameters are isotropic, then the morphology is termed as self-affine, has a fractal dimension and is generally found to be valid for various growth process such as vapour deposition of metal films, fractured surfaces and many other natural surfaces. The above mentioned formalisms that explain wetting does not consider the surface morphological parameters such as growth roughness exponent $\alpha$ and lateral correlation length $\xi$. The above formalisms also do not tell us how to modify a surface from hydrophobic to hydrophilic or vice versa by simply manipulating parameters $\alpha$ and $\xi$. To our knowledge no experimental work has been carried out to investigate the relationships between wetting and surface morphological parameters $\alpha$ and $\xi$ but some theoretical attempt has been made \cite{palas}. In this paper we will present some experimental results showing very interesting behavior of wetting depending strongly on the value of these three morphological parameters and per se we can tune the wettability of any solid surface without changing the chemical composition. 

We have used the following definitions for extracting these parameters from the data: For a given surface, if $h(x,y)$ is the height at position $(x,y)$, then one can obtain the root mean square deviation of the height which is commonly termed as rms roughness $\sigma = ({<[h(x,y) - <h(x,y)>]^2>})^{1/2}$ where the notation $<...>$ means the spatial statistical average. But, more quantitative information of the surface morphology can be extracted from the height-height correlation function $H(R)$ defined as $H(R)= <[h(x,y) - h(x',y')]^2> = 2\sigma^2-2<h(x,y)h(x',y')>$ where $R = \surd[(x-x')^2 + (y-y')^2]$ and $<h(x,y)h(x',y')> = C(R)$ is the autocorrelation function (ACF). For self-affine fractal random surface model \cite{Sinha} the ACF $C(R) = \sigma^2exp[-(R/\xi)^{2\alpha}]$ where $\xi$ and $\alpha$ are the parameters as defined earlier. In this formalism $\alpha$ describes the fractal characteristcs of the random surface and is related to Hausdorff dimension $D_f$ by $D_f=d-\alpha$, $d$ being the embedded dimension. When $\alpha = 1$, $C(R)$ is the Gaussian function, representing random surfaces of Gaussian correlation. 

Now let us introduce a ratio of rms roughness amplitude $\sigma$ (out-of-plane roughness) and in-plane roughness correlation length $\xi$ i.e., $\sigma$/$\xi$. This ratio is the measure of slope of the long wavelength undulations on the surface and is also called the rms long wavelength roughness ratio \cite{palas}. Since we know that $\alpha$ is the measure of the degree of sharp local surface irregularity then one can also define a rms local surface slope $\rho \propto \sigma/\xi^\alpha$ which represents the measure of the slope of the short wavelength undulations \cite{palas} (i.e., the slope of the sharp local irregularity). Here we shall assume that the liquid wets the crevices on the surface upon contact and this is possible only for the weak roughness surfaces \cite{palas} i.e., for $\alpha$ $\geq 0.5$ and $\sigma/\xi \ll 1$. We would like to see in this experimental study whether the contact angle has any dependence on this parameter $\rho$ which is a function of $\sigma$, $\xi$ and $\alpha$.

To obtain surfaces of different morphology, we deposited ZnO thin films layer-by-layer on glass substrates by combining a simple sol-gel and spin-coating technique. For this investigation we prepared six samples (a) 2-layered, (b) 4-layered, (c) 8-layered sample annealed at 450$^o$C and (d to f) 8-layered samples annealed at 500$^o$C, 550$^o$C and 600$^o$C respectively. The details of the sample preparation and characterisation are reported elsewhere \cite{cond-mat}. We have used Atomic Force Microscope (AFM) in the contact mode to map the morphology of the sample surface. 

\begin{figure}
\includegraphics*[width=7cm]{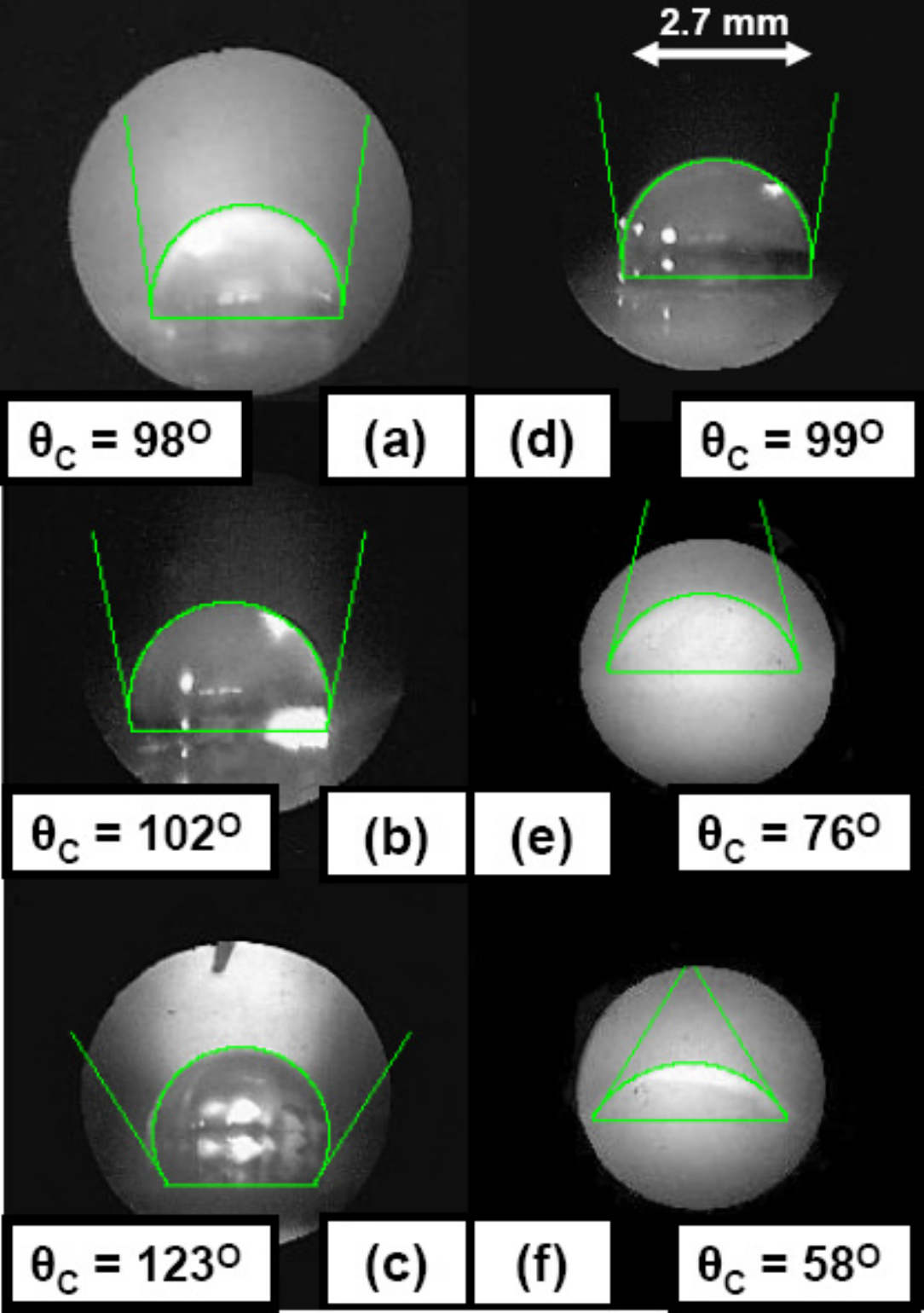}
\caption{(Colour online) CCD captured images of the 5$\mu l$ water droplets along with the contact angle measurement indicated by green lines for all the six samples. The double-sided arrow indicates the scale.}
\end{figure}

The wetting properties of these surfaces were investigated using a home made sessile drop microscope. A drop of double distilled deionized water was placed on the sample surface using a micro syringe. The image of the drop formed was analysed using ImageJ software \cite{ImageJ}. The experiment was repeated at different regions over the sample surface. In fig. 1 we show a typical captured image of the 5$\mu l$ water droplet along with the contact angle measurement for all the six samples. Two important features can be immediately observed. (1) With the increase in the number of layers for the 450$^o$C samples the contact angle progressively increases and the samples become more hydrophobic and (2) with the increase in annealing temperature the contact angle decreases and the samples become hydrophilic. To understand this behaviour we imaged the surface morphology of our samples using an AFM.

\begin{figure}
\includegraphics*[width=11cm]{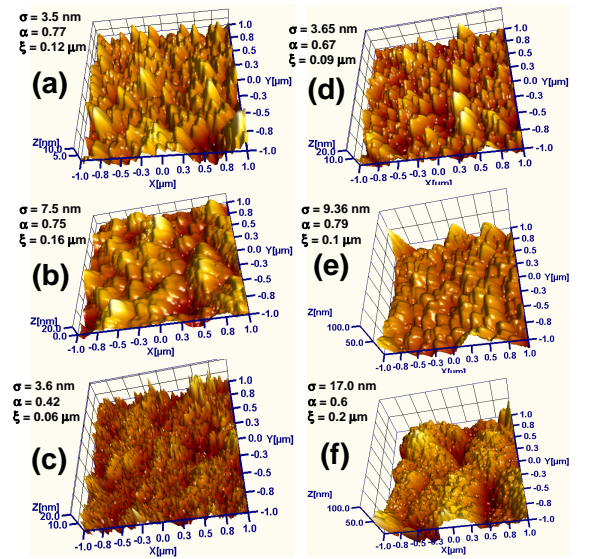}
\caption{(color online) Atomic Force Microscope (AFM) images of all the six samples, $\sigma$, $\alpha$ and $\xi$ are also indicated.}
\end{figure}

In fig. 2 we show the AFM images of all the six samples. On direct observation of the AFM images of the 450$^o$C annealed samples (samples (a) to (c), 2-layer, 4-layer and 8-layer respectively), we find that the surface morphology/asperities become more spiky with increasing number of layers, but the surface remains more or less flat. Whereas, on increasing the annealing temperature the surface does not remain flat and large surface undulations are formed (see the z-axis scale in fig. 2). The top surface morphology depends on the growth condition, and playing with these growth conditions we could obtain a set of six different surfaces having different morphological parameters. 

\begin{figure}
\includegraphics*[width=9cm]{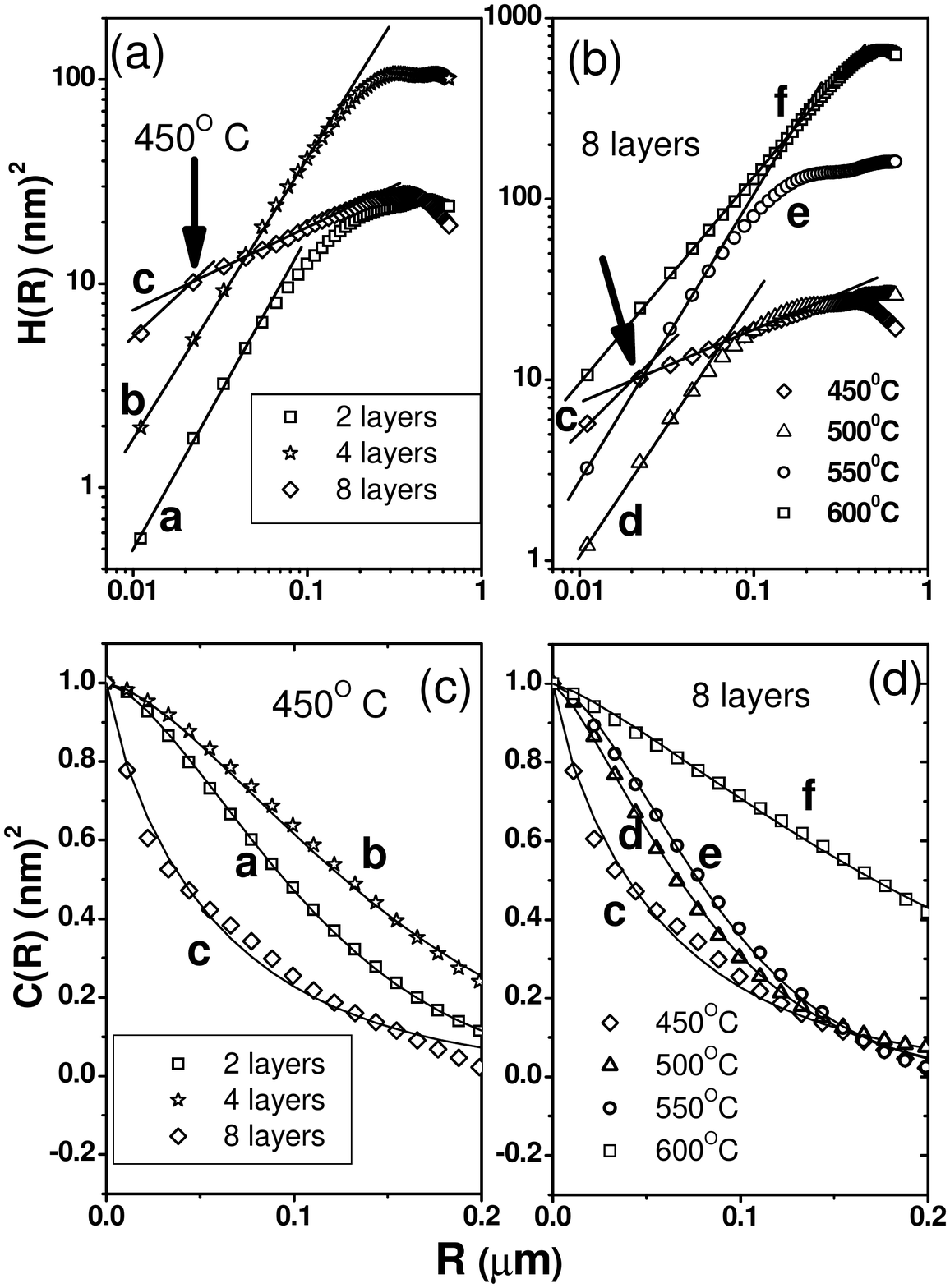}
\caption{(a,b) Height-height correlation function H(R) versus R, the value of $\alpha$ can be obtained from the fit. (c,d) we show the ACF C(R) as a function of R and the solid line shows the fit to obtain the parameter $\xi$. The arrows indicate the cross over of the roughness exponent around R $\sim$ 0.02 $\mu$m}
\end{figure}

\begin{figure}
\includegraphics*[width=8cm]{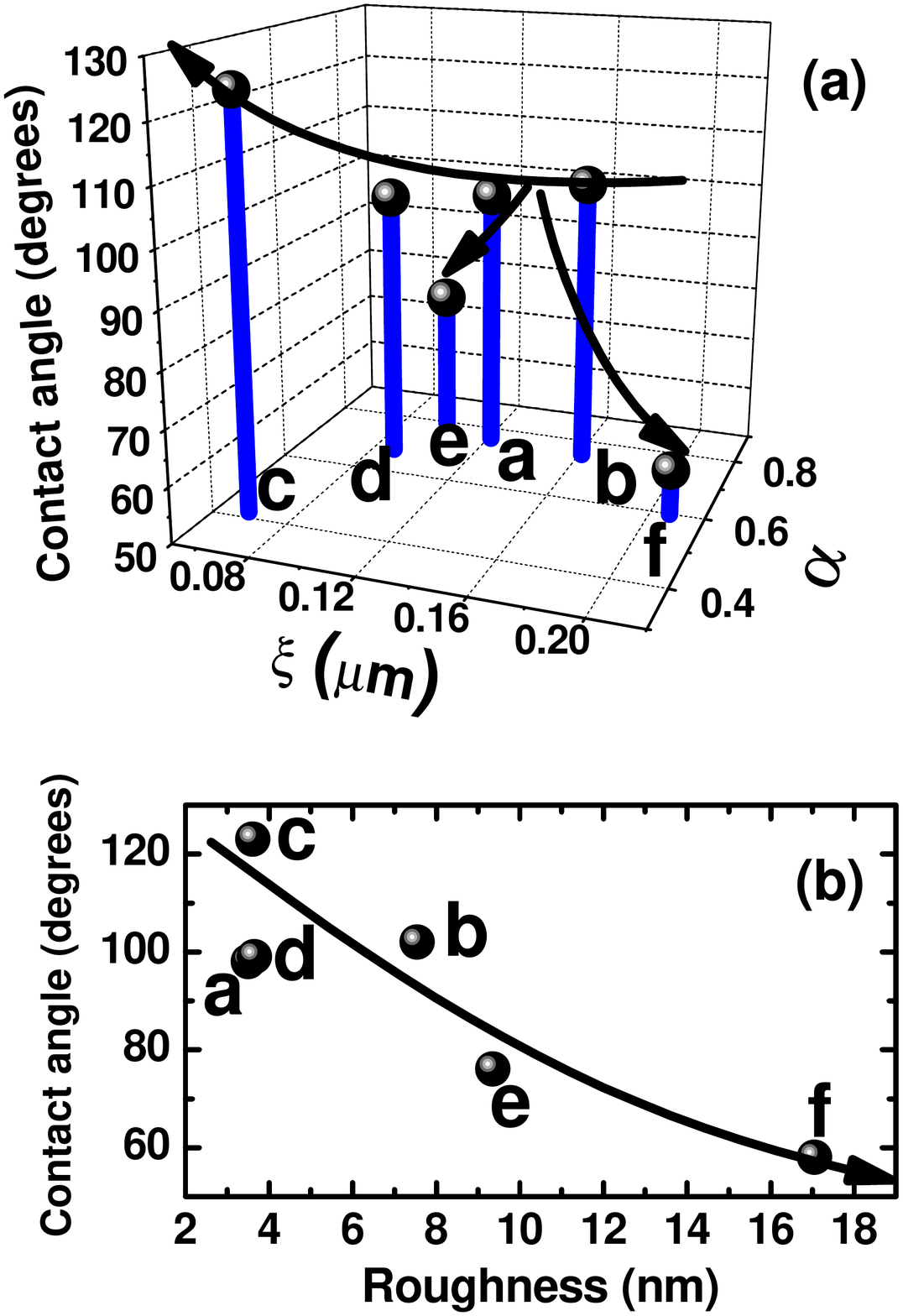}
\caption{(Color online)(a) 3D plot of contact angle versus in-plane correlation length $\xi$ and the roughness exponent $\alpha$. (b) contact angle versus rms surface roughness
The error in the contact angle is represented by the size of the data symbols (dots).}
\end{figure}

To obtain the quantitative information of the surface morphology we have extracted all the three parameters i.e., (1) $\sigma$, (2) $\xi$ and (3) $\alpha$ using the formalism described earlier. In figs. 3(a,b) we have plotted the height-height correlation function H(R) versus R and since $H(R) \propto R^{2\alpha}$ when $R << \xi$ the value of $\alpha$ can be obtained from the fit. We also observe that H(R) tends to 2$\sigma^2$ when $R >> \xi$ as expected for the self-affine surface morphology. It is interesting to note that for sample (c) [450$^o$C, 8-layers] we observe a crossover of the roughness exponent around $R$ $\sim$ 0.02 $\mu m$ indicating that the surface morphology has an hierarchical structure. It has been predicted theoretically that this sort of heirarchial structure will exhibit dewetting property of the solid surface \cite{Herminghaus}. We could also obtain the value of the lateral correlation length $\xi$ using the above mentioned formalism and by using the value of $\alpha$ obtained from H(R). In figs. 3(c,d) we show the ACF C(R) as a function of R and the solid line shows the fit from which the parameter $\xi$ has been obtained. We observe that for sample (c) the surface appears spiky (i.e., sharp local irregularities) and flat, leading to low $\alpha$ and $\xi$ similar to the morphology proposed by Cassie-Baxter. This leads to a high contact angle of the liquid drop. We see that the sample (c) does not obey the weak surface roughness criteria as discussed above and hence this sample (c) will not scale with the rms local surface slope $\rho$. For sample (f) even though the roughness exponent $\alpha$ is low indicating large local slope of the sharp irregularities, the rms roughness amplitude $\sigma$ and the in-plane lateral correlation length $\xi$ is larger indicating large undulations on the surface and the contact angle of the water drop is observed to be the lowest. This aspect will be addressed further below. In fig. 4(a) we show a 3D plot of the contact angle versus in-plane correlation length $\xi$ and the roughness exponent $\alpha$. We clearly observe that the contact angle is a non-monotonic function of $\alpha$ and $\xi$. In fig. 4(b) we observe another interesting feature i.e., we observe that as the rms surface roughness $\sigma$ increases the contact angle decreases.  Now we would like to investigate whether there exists any closed functional form exits between the contact angle and all these surface morphological parameters $\alpha$, $\sigma$ and $\xi$.

From the above experimental observation we can infer that one can tune the contact angle by controlling or tuning the surface morphological parameters. Large contact angle (hydrophobic) for sample (c) could be explained by invoking Cassie-Baxter formalism, but how do we explain the hydrophilic nature obtained for sample (f) having moderately low roughness exponent $\alpha$ but relatively large lateral correlation length $\xi$ because of prominent undulations (mounds/peaks and troughs/valleys) on the sample surface (see fig.~2(f)) leading to topography driven wetting? Seperation between the peaks and valleys (the undulations) is the lateral correlation length $\xi$. For sample (f) the asperities are riding over the undulations. This can be observed directly from the AFM image and was also observed from the fast Fourier transform of the image. If undulations are introduced on the surface then the liquid will follow the undulations and will flow/spill-over into the trough/valleys (crevices) unlike in the Cassie-Baxter case where the liquid cannot break the liquid film and fill the air-pockets between the dense asperities on a flat substrate. The spill-over of liquid into the adjacent trough/valleys can happen because large fraction of these valleys will be connected because of large surface rms roughness (as can be seen from the z-scale of fig. 2(f)). The penetration of the liquid to this adjacent troughs/valleys is very similar to the model (penetration model) proposed by Ishino et al \cite{Ishino}. Large value of $\sigma$ can only give rise to the formation of large troughs/valleys such that the liquid can spill-over into these. For the case of sample (f) eventhough the value of $\alpha$ is low, we obtain from our analysis a large value of the slope of the long wavelength undulation on the surface ($\sigma/\xi$), thus preventing the surface to go to the Cassie-Baxter limit. These large undulations help in transforming the hydrobhobic surface to a hydrophilic surface. Thus, our experimental results indicate that all the three surface morphological parameters $\sigma$, $\xi$ and $\alpha$ are important to determine the wetting property of the solid surface.

\begin{figure}
\includegraphics*[width=8cm]{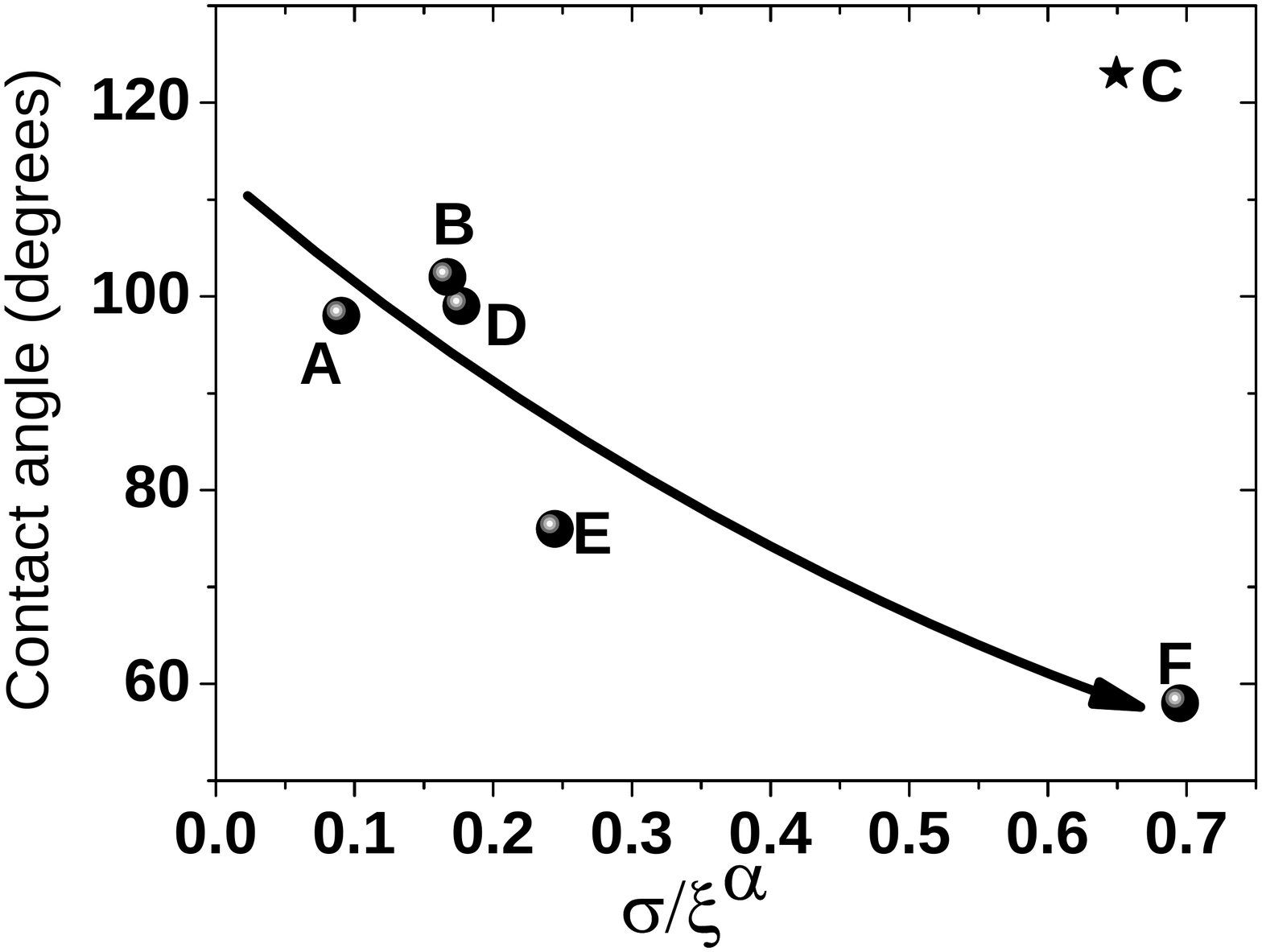}
\caption{Contact angle verses $\rho$ ($\propto \sigma/\xi^\alpha$) the rms local surface slope. The surface of Sample C is in the Cassie-Baxter limit.}
\end{figure}

Next, we tried to see if the contact angle has any dependence on these three surface morphological parameters $\sigma$, $\xi$ and $\alpha$. In fig.~5 we have plotted the contact angle versus rms local surface slope $\rho$. We observe that the contact angle has a monotonic dependence on $\rho$ ie., the contact angle decreases with increasing value of $\rho$. The sample (c) is shown as an isolated point since it is a true Cassie-Baxter surface. From fig.~5 it appears that the dependence of contact angle on $\rho$ is valid only for the cases where the liquid wets the crevices of the surface upon contact.

To summarize, in this report we have characterized the surface morphology of sol-gel grown ZnO multilayer films using atomic force microscopy and obtained the surface morphological parameters. These parameters were correlated to the wetting property (contact angle) of the liquid over the sample surface. We found that the contact angle monotonically depends on the rms local surface slope $\rho$ for a self-affine fractal surface when the liquid wets the crevices of the sample surface. $\rho$ is a function of all the three surface morphological parameters $\rho(\sigma,\xi,\alpha)$. We could infer from our experimental results that we can tune the wetting property (of self-affine fractal surfaces) by merely changing the morphological parameters of the surface.

Acknowledgement: The authors thank P. Misra for his help in obtaining the AFM images. S. Sarkar and S. Patra would like to thank SINP for giving opportunity to work on this project.


\begin{thebibliography}{99}
\bibitem{ImageJ} ImageJ is a public domain, Java-based image processing program developed at the National Institutes of Health.
\end{thebibliography}
\end{document}